\documentclass[journal]{IEEEtran}
\usepackage{cite}
\usepackage{amsmath,amssymb,amsfonts}
\usepackage{algorithmic}
\usepackage{graphicx}
\usepackage{textcomp}
\usepackage[colorlinks=true,linkcolor=black,citecolor=black,urlcolor=blue]{hyperref}
\usepackage{mhchem}
\def\BibTeX{{\rm B\kern-.05em{\sc i\kern-.025em b}\kern-.08em
    T\kern-.1667em\lower.7ex\hbox{E}\kern-.125emX}}
\newcommand{\journalname}{}
\begin{document}
\title{A Magnon-Based Electric Field Controlled Magnetoelectric Device for Energy-Efficient Logic-in-Memory}
\author{Rongqing Cong*, \IEEEmembership{Graduate Student Member, IEEE}, Sajid Husain*, \IEEEmembership{Member, IEEE},  
Yumin Su, \IEEEmembership{Graduate Student Member, IEEE},
Sasikanth Manipatruni, \IEEEmembership{Senior Member, IEEE},
Naveed Ahmed, \IEEEmembership{Student Member, IEEE}, Dmitri E. Nikonov, \IEEEmembership{Senior Member, IEEE},
Ramamoorthy Ramesh, \IEEEmembership{Member, IEEE},
Kaiyuan Yang, \IEEEmembership{Senior Member, IEEE},
Zhi Jackie Yao, \IEEEmembership{Member, IEEE}
\vspace{-0.7cm}
\thanks{*These authors contributed equally. Rongqing Cong, Yumin Su, Naveed Ahmed and Kaiyuan Yang are with the Department of Electrical and Computer Engineering, Rice University, Houston, TX 77005, USA. Sajid Husain and Ramamoorthy Ramesh are with the Department of Materials Science and Engineering, University of California, Berkeley, Berkeley, CA 94720, USA. Sasikanth Manipatruni and Dmitri E. Nikonov are with Kepler Computing Inc., Hillsboro, Oregon 97006, USA. Zhi (Jackie) Yao  is with the Applied Mathematics and Computational Research Division, Lawrence Berkeley National Laboratory, USA, Berkeley, CA 94720, USA (e-mail: jackie\_zhiyao@lbl.gov).}}
\maketitle

\begin{abstract}
We demonstrate a non-volatile magnetoelectric magnonic memory (MEMM) that enables fully electrical write/read via direct magnon-driven sensing in an insulating antiferromagnet. A fabricated SrIrO$_3$/La-BiFeO$_3$/SrIrO$_3$ trilayer exhibits sub-100\,ps switching, a remnant polarization of 20\,$\mu$C/cm$^2$, and a readout voltage contrast close to 1\,mV between high- and low-resistance states. To connect device physics to circuit behavior, we develop and experimentally validate a compact circuit model that captures spin Hall injection and spin transport. Simulations with optimized material parameters predict output voltages $>100$\,mV, enabling cascading without external amplification. Using this framework, we design MEMM-based memory and logic blocks, including a 1T1R array, two inverter implementations (complementary two-device and single-device), and a three-input majority gate, and evaluate deep-pipelined operation. The model projects switching energies down to 1\,aJ per operation and logic propagation delays of 30-60\,ps, indicating MEMM as a promising platform for energy-constrained, high-throughput computing.
\end{abstract}


\section{Introduction}
\normalcolor

The scaling of CMOS energy efficiency is fundamentally limited by the Boltzmann tyranny ($\sim$0.4V thermal voltage barrier) and the slowdown of Moore’s law, intensifying interest in spintronic devices for their low-power and non-volatility. By enabling in-memory computing, these devices have also shown potential to reduce data transfer energy at the architectural level. However, most spintronic memories, including spin-transfer-torque (STT) and spin-orbit-torque (SOT) magnetoresistive random access memory (MRAM), still rely on current-driven magnetic switching, resulting in substantial energy loss due to Joule heating\cite{liao_evaluating_2022, worledge_spin-transfer_2024}.

\begin{figure}
\centerline{\includegraphics[width=\columnwidth]{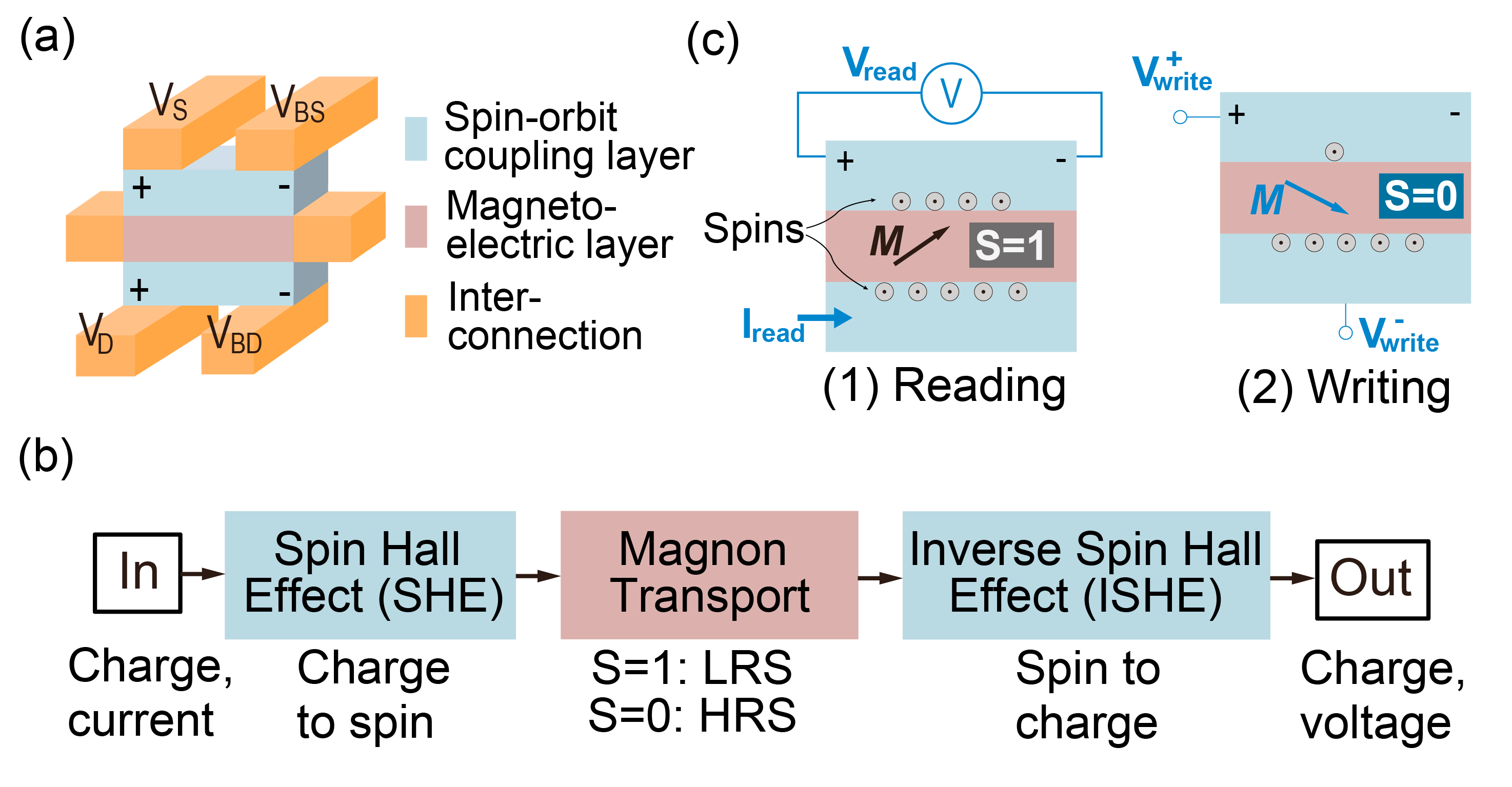}}
\caption{Working principles of the trilayer MEMM device: (a) Device structure, composing of SO coupling layer, ME layer and interconnections; (b) Operating mechanism. The device state is defined as S=1 for the LRS and S=0 for the HRS; (c)The device operates in two modes: reading the stored state and writing a new state by applying switch voltages.\vspace{-1.0em}}
\label{fig:structure}
\end{figure}

To overcome the energy‐loss limitations of charge‐driven spintronics, recent work has been relying on magnons (quantized spin waves) that carry angular momentum without charge motion, thus avoiding Joule heating. Experiments demonstrate that magnons propagate with minimal dissipation in synthetic antiferromagnets (AFM)\cite{millo_unidirectionality_2023} and that their populations can be detected electrically via thermal spin‐wave noise \cite{devolder_measuring_2022}, confirming their promise for low-energy-loss information transfer. However, conventional magnonic devices are volatile, as the information is encoded in the phase and amplitude of the spin waves \cite{balinskyy_magnonic_2024}. Voltage-controlled magnetoelectric spin-orbit (MESO) transducers, which use ferromagnetic (FM) composites and magnetoelectric (ME) thin films \cite{manipatruni_scalable_2019, vaz_functional_2021}, address volatility through the ferroelectricity of the ME layer, allowing non-volatile memory and logic operations. Moreover, electric-field-driven spin-wave excitation and detection make these devices compatible with CMOS platforms \cite{liao_evaluating_2022,narla_cross-layer_2025}. Nevertheless, MESO devices suffer from limited endurance at the FM-ME interface, which restricts their practical use in logic circuits. 

Building on these advances, we introduce a magnetoelectric magnonic memory device (MEMM) that leverages (1) electric‐field control of magnetization through an ME material and (2) electron‐free signal transport through magnon propagation in an insulating AFM. Fig.~\ref{fig:structure}(a) illustrates the trilayer device structure, consisting of an ME layer, e.g. \ce{BiFeO3} (BFO) and \ce{La-BiFeO3} (LBFO), sandwiched between two spin-orbit (SO) layers, e.g. SrIrO$_3$ (SIO). The operating principle, shown in Fig.~\ref{fig:structure}(b), is based on voltage-controlled switching of the ferroelectric polarization in (L)BFO, which modulates the magnon transport resistance between the high-resistance state (HRS, S=0) and the low-resistance state (LRS, S=1). As shown in Fig.~\ref{fig:structure}(c), the device operates in two modes: reading the stored state through spin current and writing a new state by applying switching voltages across the ME layer. The rest of the manuscript is organized as follows: In Section II, we experimentally demonstrated both the switch and the read functionality of the device. In Section III, we developed and validated a circuit model against experimental data, showing that with optimized parameters, the device can produce over 100 mV of output voltage, sufficient to drive subsequent MEMM cells. In Section IV, utilizing this scalable behavior, we designed memory arrays and logic circuits that exhibit the potential for low-power, high-density in-memory computing. Benefiting from its compact structure, we also showed the MEMM switches in under 100ps. 

\begin{figure}[!t]
\centerline{\includegraphics[width=\columnwidth]{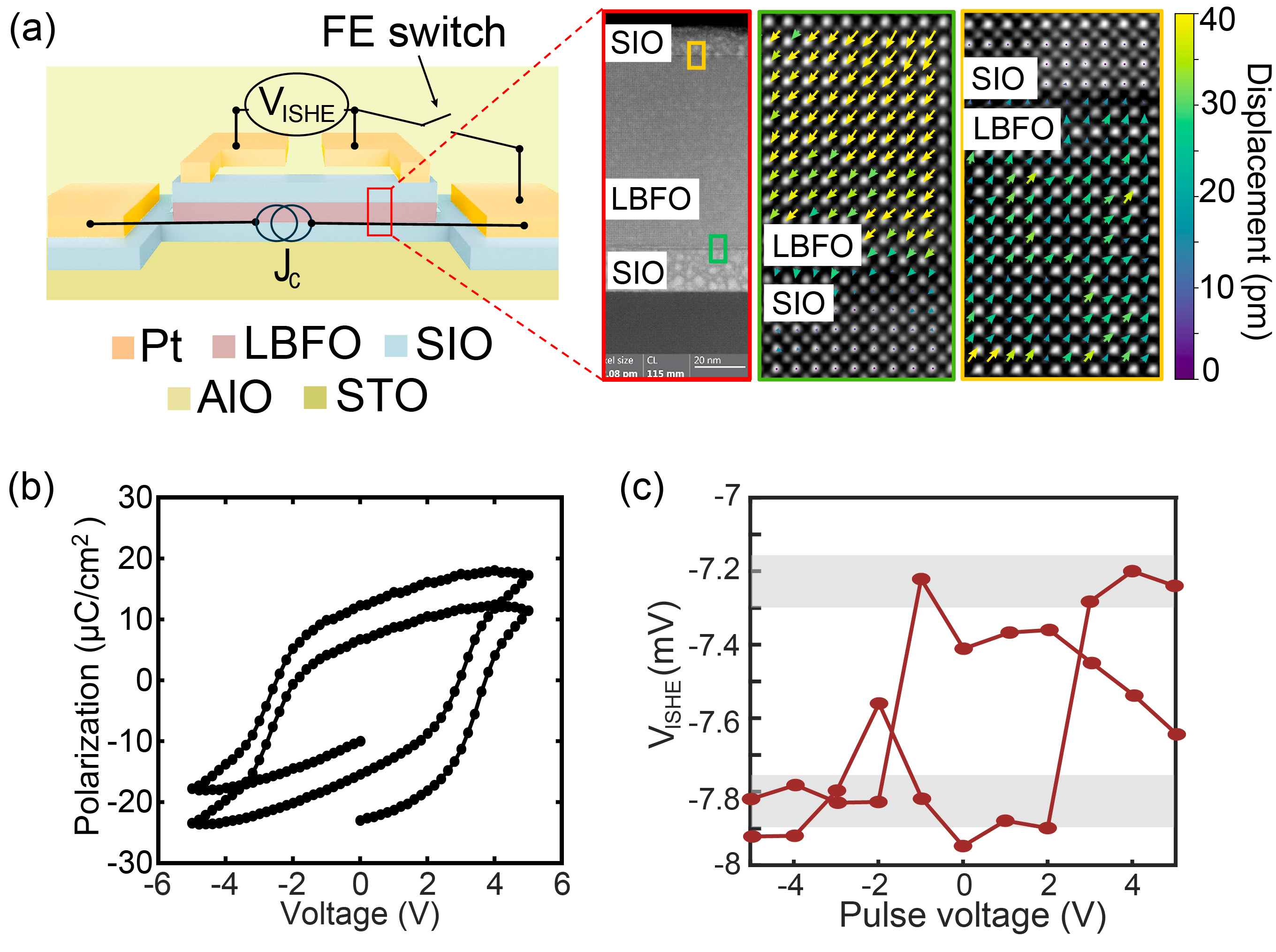}}
\caption{(a) Schematic of a SrIrO$_3$ (SIO)/La-BiFeO$_3$ (LBFO)/SrIrO$_3$ (SIO) trilayer device and high resolution cross-sectional TEM image and vector polarization mapping. (b) Ferroelectric (polarization-electric field hysteresis) response of the LBFO from the SIO/LBFO/SIO trilayer. (c) The first harmonic lock-in measurement of output voltage $V_\text{ISHE}$ as a function of switch voltage. }
\label{fig:device-demo}
\end{figure}

\section{Device Demonstration}
\normalcolor

Fig.~\ref{fig:device-demo}(a) illustrates the trilayer MEMM device schematic alongside its corresponding switching and readout operations. The structure was epitaxially grown on a \ce{SrTiO_3} substrate via pulsed laser deposition. 
High-angle annular dark-field scanning transmission electron microscopy (HAADF-STEM) confirms layer-by-layer epitaxy with sharp \ce{SIO}/\ce{LBFO} interfaces (right panel). 
This atomic-scale uniformity is critical, as it minimizes spin scattering at the interfaces to ensure efficient spin-current injection and magnon transduction.

Device functionality includes two distinct phases: polarization switching and non-volatile state readout.
1. Switching: Voltage pulses applied across the top and bottom SIO electrodes reverse the ferroelectric polarization of the intermediate LBFO layer. 
This reorientation sets the LBFO's magnon transport into either LRS or HRS. Fig.~\ref{fig:device-demo}(b) presents the corresponding ferroelectric hysteresis loop, exhibiting a well-defined square shape with a remnant polarization of $\sim$20 $\mu$C/cm$^2$. This confirms robust, bistable polarization suitable for non-volatile memory. 
2. Readout: The stored polarization state is detected electrically via magnon transport. A constant charge current ($J_C$) in the bottom SIO layer generates a vertical spin current via the spin Hall effect (SHE). This spin accumulation at the interface excites magnons within the LBFO. After propagating through the antiferromagnetic insulator, the magnons are reconverted into a detectable transverse voltage ($V_\textrm{ISHE}$) in the top SIO layer via the inverse spin Hall effect (ISHE).
Fig.~\ref{fig:device-demo}(c) shows the non-volatile readout signal, where $V_\textrm{ISHE}$ is plotted against the switching voltage. The first-harmonic lock-in measurement reveals a clear hysteresis, directly correlating the ferroelectric state with magnon conductance. Two distinct output voltage levels ($\Delta V \approx 1$ mV) correspond to the opposite polarization orientations, with switching thresholds aligned to the coercive fields in Fig.~\ref{fig:device-demo}(b). The following sections discuss pathways to enhance this on/off ratio for integrated memory and logic circuits.

\begin{figure}[t]
    \centering
    \includegraphics[width=\linewidth]{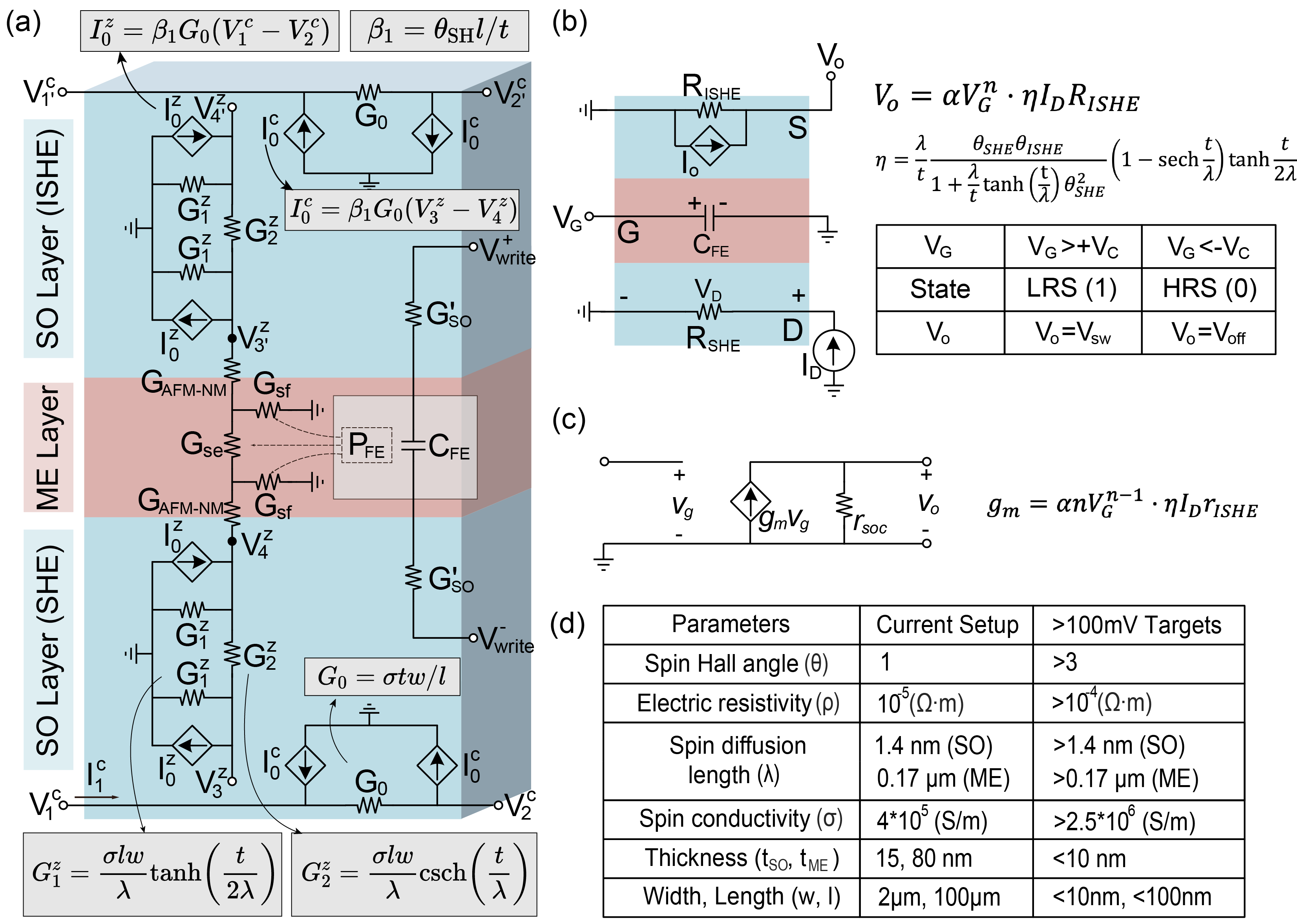}
    \caption{(a) An equivalent spin circuit model illustrating the processes from bottom to top: spin Hall effect, magnon transport, and inverse spin Hall effect. (b) Simplified circuit model showing control signal $V_G$, supply current $I_D$, and output voltage $V_o$. The equation on the right describes the relationship between output, control, and supply, where $\eta$ is a material-dependent parameter and $\alpha V_G^n$ is curve-fitted from the hysteresis loop, exhibiting a sharp increase near the threshold voltage $V_C$. (c) Equivalent small-signal circuit model. (d) Demonstrated device parameters and target values required to achieve output voltages of 100mV and above.\vspace{-1.0em}} 
    \label{fig:model}
\end{figure}

\vspace{-1.0em}
\section{Device Modeling and Optimization}
\normalcolor

\subsection{Spin Circuit Model}

We developed an equivalent spin circuit model (Fig.~\ref{fig:model}a) for the MEMM device using vector spin circuit theory~\cite{hong_spin_2016}, enabling SPICE-based analysis of its electrical behavior. For simplicity, all three layers are assumed to have identical width $w$ and length $l$, with adjustable thicknesses $t_\text{ME}$ (magnetoelectric layer) and $t_\text{SO}$ (spin-orbit layers). Material parameters include spin Hall angle $\theta_\text{SH}$, spin diffusion length $\lambda$, and charge conductivity $\sigma$.
The model employs a nodal representation where each terminal $i$ has charge ($\eta = c$) or spin ($\eta = x, y, z$) voltages $V_i^\eta$ and currents $I_i^\eta$. The spin voltage and current are vectors, but we consider only the $z$-component for this device geometry.
Charge-spin interconversion is modeled by controlled current sources. At the bottom spin-orbit layer, SHE generates a spin current proportional to the transverse charge voltage:
\begin{equation}
I_0^z = \beta_1 G_0 (V_1^c - V_2^c),
\label{eq:current_sources}
\end{equation}
where $G_0 = \sigma t w / l$ is the layer's charge conductance. At the top layer, ISHE produces a charge current from the spin voltage:
\begin{equation}
I_0^c = \beta_1 G_0 (V_3^z - V_4^z).
\label{eq:current_sources}
\end{equation}
The coupling coefficient $\beta_1 = \theta_\text{SH} l / t$ quantifies the conversion efficiency.
Spin transport within each layer is governed by diffusion equations, yielding spin conductances:
\begin{equation}
G_1^z = \frac{\sigma l w}{\lambda} \tanh\left(\frac{t}{2\lambda}\right), 
G_2^z = \frac{\sigma l w}{\lambda} \text{csch}\left(\frac{t}{2\lambda}\right).
\label{eq:conductances}
\end{equation}

The magnetoelectric layer's spin conductances (spin-elastic $G_\text{se}$ and spin-flip $G_\text{sf}$) are modulated by ferroelectric polarization. For polarization state `1', $G_\text{se}$ is high and $G_\text{sf}$ is low, allowing nearly unimpeded spin current flow with minimal voltage drop between $V_4^z$ and $V_{3'}^z$. Conversely, for state `0', $G_\text{se}$ is low and $G_\text{sf}$ is high, significantly attenuating the spin current. This non-volatile conductance switching enables memory functionality.

In operation, a charge current injected into the bottom layer creates a spin current via SHE. This spin current propagates through the ME layer, with its amplitude controlled by the ferroelectric state. The transmitted spin current is then converted to a measurable charge voltage at the top layer via ISHE, providing electrical readout of the stored state.

\begin{figure}[h]
    \centering
    \includegraphics[width=1\linewidth]{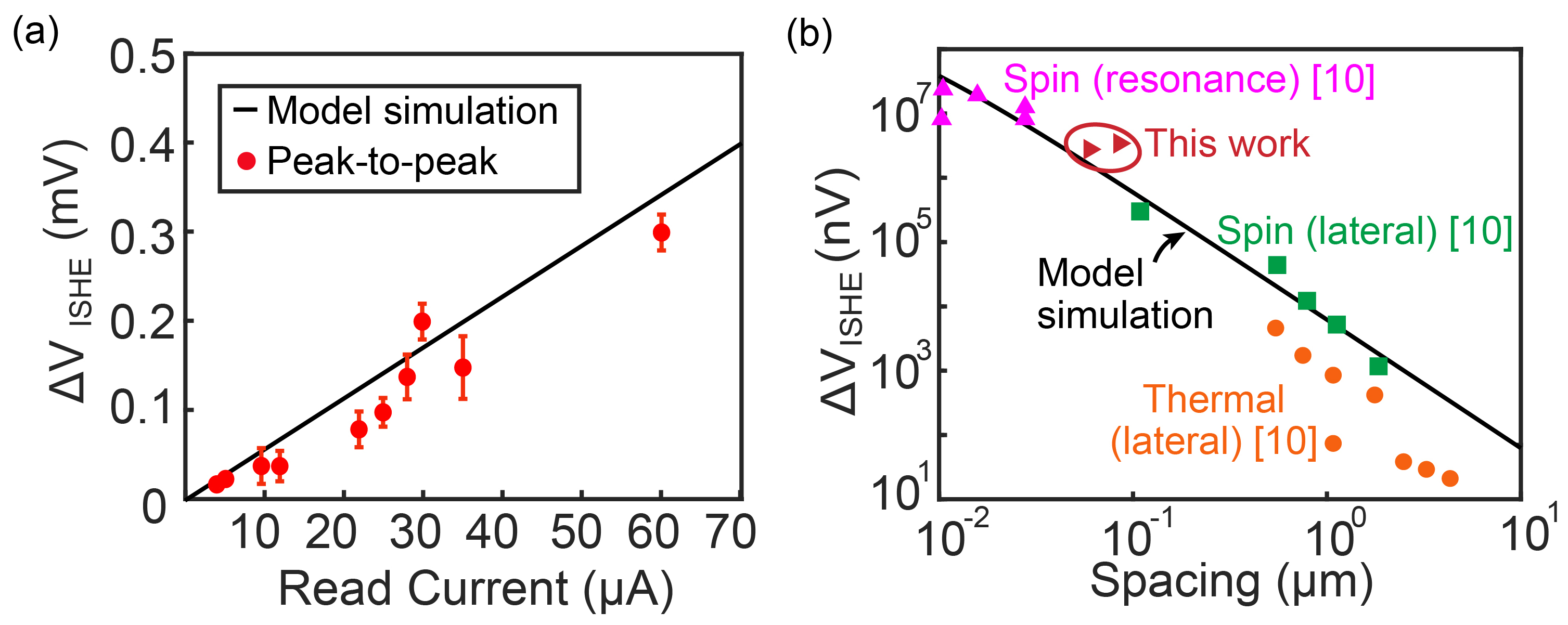}
    \caption{(a) Comparison between simulation results and experimental data, extracted from the peak-to-peak voltage in the hysteresis loop. (b) Comparison of simulation results with previous studies \cite{huang_manipulating_2024} and our experimental data, revealing a consistent log-log relationship of voltage and spacing ($t_\text{SO}$, the thickness of top and bottom layer).}
    \label{fig:exper and simulation}
\end{figure}

\begin{figure}
    \centering
    \includegraphics[width=1\linewidth]{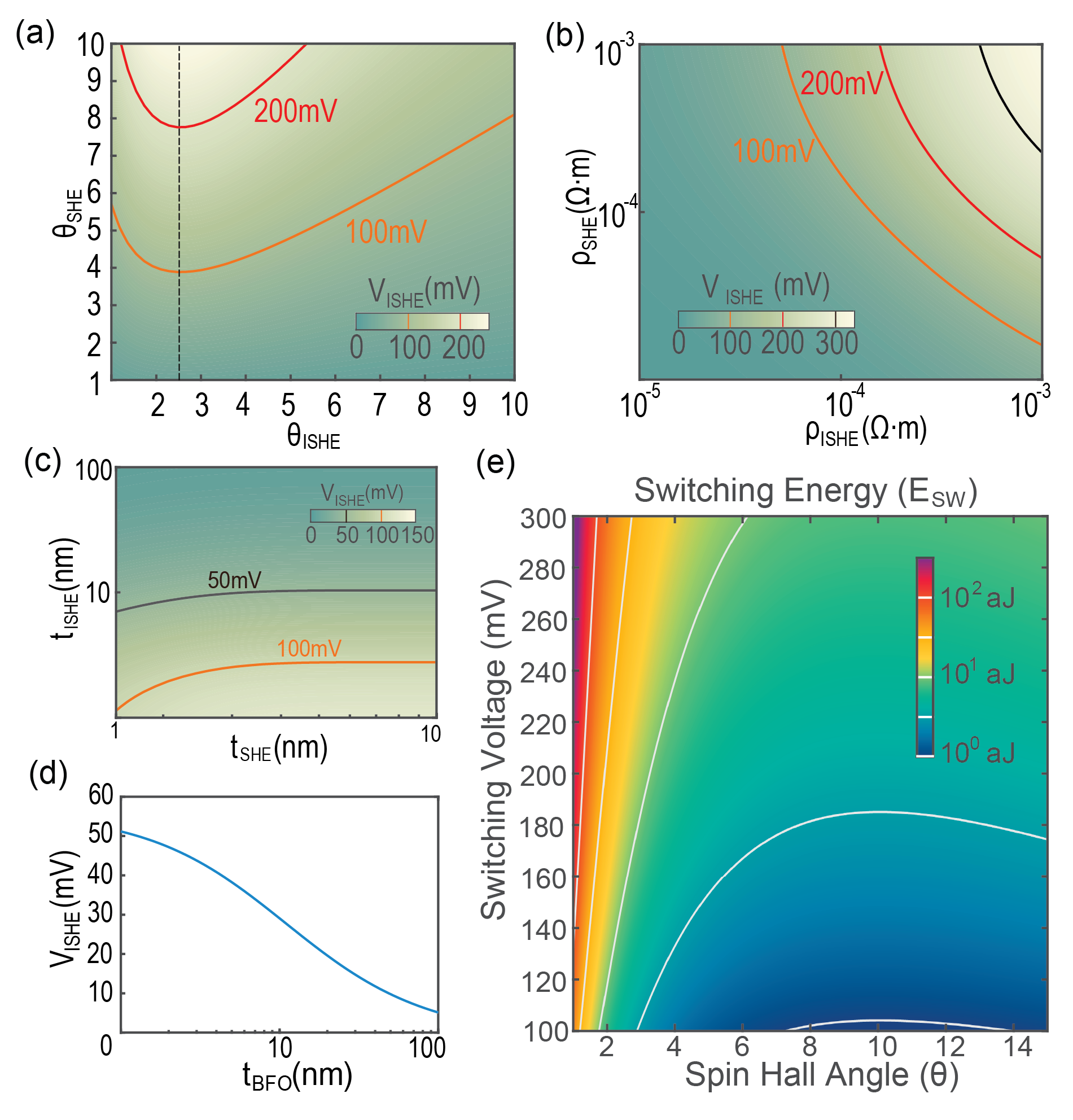}
    \caption{Design space exploration for MEMM device optimization.
(a) Output voltage contours as functions of spin Hall angles for the bottom ($\theta_{\text{SHE}}$) and top ($\theta_{\text{ISHE}}$) layers.
(b) Output voltage contours as functions of electrical resistivities for the bottom ($\rho_{\text{SHE}}$) and top ($\rho_{\text{ISHE}}$) layers.
(c) Impact of spin-orbit layer thicknesses ($t_{\text{SHE}}$, $t_{\text{ISHE}}$) on output voltage, illustrating scalability.
(d) Output voltage as a function of magnetoelectric layer thickness ($t_{\text{BFO}}$).
(e) Simulated switching energy from our model, demonstrating the potential for ultra‑low‑energy operation in the attojoule (1 aJ) range.
\vspace{-1.5em}}
    \label{fig:design-space}. 
\end{figure}

\vspace{-1.0em}
\subsection{Single Device Simulation Results}

To validate our modeling approach, we compared simulation results with experimental data (Fig.~\ref{fig:exper and simulation}). 
The model verification demonstrates excellent agreement: Fig.~\ref{fig:exper and simulation}(a) shows a linear relationship between the read current and the output on/off voltage difference $\Delta V_{\text{ISHE}}$, resembling the characteristic linear-region behavior of CMOS devices. 
Fig.~\ref{fig:exper and simulation}(b) confirms promising device scalability, demonstrating that an input-output spacing below 100~nm can achieve an output voltage difference close to 100~mV. These results were obtained using the parameters listed in the table of Fig.~\ref{fig:model}, which represent our current experimental conditions.

We further utilized this validated model to explore the design space by sweeping geometric and electrical parameters to map achievable device performance. As illustrated in Fig.~\ref{fig:model}(b), optimizing key parameters -- specifically increasing the spin Hall angle and reducing the switching voltage -- enables switching energies to reach the attojoule regime. The final column of the table in Fig.~\ref{fig:model} presents our targeted parameter specifications for achieving ideal low-power switching performance. To detail these specifications, we present comprehensive parametric sweeps across key geometric and material properties in Fig.~\ref{fig:design-space}.

Fig.~\ref{fig:design-space}(a) illustrates the effect of the spin Hall angles of the bottom and top spin-orbit layers, $\theta_{\text{SHE}}$ and $\theta_{\text{ISHE}}$, on the achievable output voltage $V_{\text{ISHE}}$, indicated by the color contours. The analysis reveals that increasing both spin Hall angles enables significantly higher output voltages, with optimal operating points occurring when $\theta_{\text{ISHE}}$ is approximately 2.5. This suggests that materials with large spin Hall angles are essential for cascadable logic operation without requiring additional amplification stages.
Fig.~\ref{fig:design-space}(b) examines how the electrical resistivities of both spin-orbit layers affect device performance. The log-log plot demonstrates that maintaining output voltages above 100~mV requires careful optimization of both $\rho_{\text{SHE}}$ and $\rho_{\text{ISHE}}$, revealing a clear trade-off between material conductivity and signal strength. Lower resistivity values generally correlate with higher output voltages due to reduced Joule heating losses and more efficient current flow, indicating that highly conductive spin Hall materials are preferred for energy-efficient operation.
Fig.~\ref{fig:design-space}(c) investigates scalability by analyzing the impact of spin-orbit layer thicknesses $t_{\text{SHE}}$ and $t_{\text{ISHE}}$ on output voltage. The results show that output voltage decreases logarithmically with increasing layer thickness, suggesting that thinner spin-orbit layers (in the range of 1-10~nm) provide higher voltage outputs due to enhanced current density and more efficient spin-charge conversion. This scaling behavior provides important guidance for device miniaturization while maintaining adequate signal levels.
Fig.~\ref{fig:design-space}(d) examines the relationship between ME layer thickness $t_{\text{BFO}}$ and output voltage. The monotonically decreasing trend indicates that thinner LBFO layers yield higher output voltages, likely due to reduced magnon scattering and shorter transport distances. However, practical constraints such as ferroelectric stability and switching reliability impose a lower bound on the magnetoelectric layer thickness. 
Finally, Fig.~\ref{fig:design-space}(e) demonstrates that sub-attojoule ($<1~aJ$) switching energy can be achieved when the spin Hall angle exceeds 3 (under the symmetric condition $\theta_{\text{SHE}} = \theta_{\text{ISHE}}$), provided the switching voltage is maintained above a practical threshold.
These parametric studies collectively establish clear, albeit trade-off-sensitive, design rules for making high-performance MEMM devices that minimize energy consumption while maximizing output signal strength, which are key requirements for cascaded logic operations.

\vspace{-0.5em}
\section{Memory and Logic Circuit Design}

Building upon the working mechanism of MEMM devices introduced previously, this section explores their implementation in practical memory and logic circuits, leveraging their distinct spin conductivity states modulated by ferroelectric polarization.

\vspace{-0.5em}
\subsection{Memory Array}

\begin{figure}[t]
    \centering
    \includegraphics[width=1\linewidth]{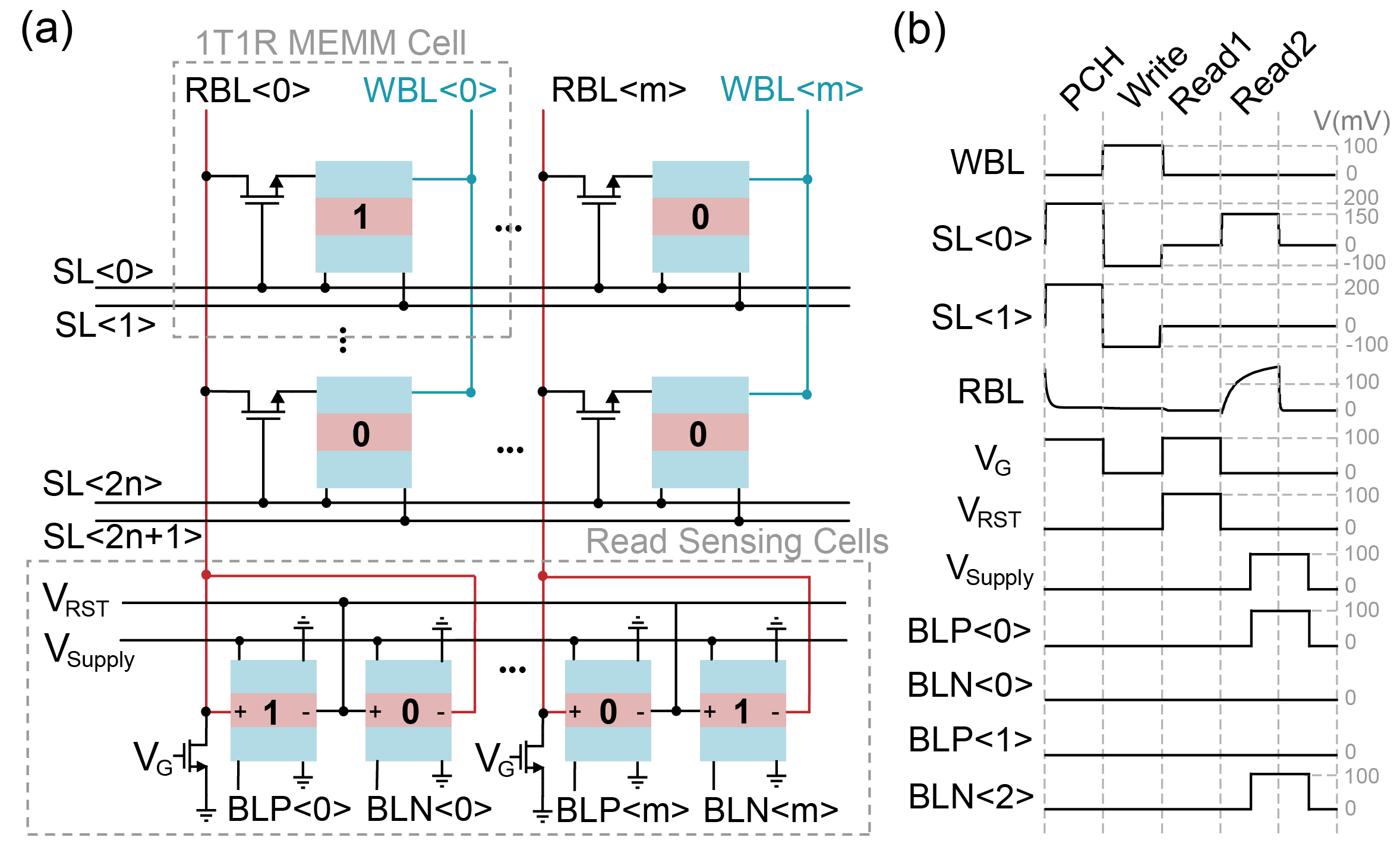}
    \caption{(a) 1T1R memory array design using MEMMs. (b) Writing and reading operating waveforms. In this example, we assume the switching voltage $V_T=200mV$ and reading voltage $V_R=150mV$. }
    \label{fig:memory}
\end{figure}

Leveraging ferroelectric polarization-modulated spin conductivity, MEMM devices are promising candidates for low-power memory applications. We designed a compact 1T1R (one transistor, one resistor) memory array architecture as shown in Fig.~\ref{fig:memory}a, where each MEMM cell is accessed through a single transistor controlled by word bitlines (WBL) and read bitlines (RBL).

The memory operation consists of three distinct phases illustrated in the timing waveforms of Fig.~\ref{fig:memory}b. In the pre-charge phase (PCH), all MEMM cells across the array are uniformly initialized to logic state `0' by biasing the sense lines (SL) to the switching threshold voltage $V_T = 200\text{mV}$. This global initialization ensures a consistent starting state for subsequent operations while the supply voltage $V_{\text{Supply}}$ remains at ground potential.

During the write phase, data is programmed row-by-row by biasing the two sense lines of the selected row (SL$\langle$2i$\rangle$ and SL$\langle$2i+1$\rangle$) to $-V_T/2 = -100\text{mV}$ while simultaneously driving the target write bitline (WBL$\langle$j$\rangle$) to $+V_T/2 = +100\text{mV}$. This differential addressing applies the full switching voltage $V_T$ (200 mV) across only the targeted MEMM cell, while non-selected cells in the same row experience only $V_T/2$, preventing unintended writes and ensuring data integrity.

The readout operation utilizes a two-phase, non-destructive sensing scheme. It exploits the difference in spin conductivity between the logic `0' and `1' states.
Phase 1 (Reset): The dedicated sensing cells are first reset to a known reference state by biasing their supply lines.
Phase 2 (Read): The selected sense line, SL$\langle$2i$\rangle$, is driven with a read voltage $V_R = 150\text{mV}$. This voltage is deliberately kept lower than the write threshold to prevent disturbing the stored state. The cell's magnetic state modulates the spin current flowing through it, which in turn perturbs the voltage on the corresponding read bitline (RBL).
Signal Conversion: This perturbation on the RBL is detected by the sensing cells, which convert it into a strong, differential voltage signal between the complementary bitline pair (BLP and BLN). The polarity and magnitude of this differential signal directly correspond to the stored logic value.
A key benefit of this approach is that the differential output signal is robust enough to interface directly with downstream logic gates, eliminating the need for separate sense amplifiers. This enables tight integration of memory and logic within the same array. The underlying 1T1R architecture, combined with this differential sensing method, ensures the design is scalable to large, high-density memory applications.

\begin{figure}
    \centering
    \includegraphics[width=1\linewidth]{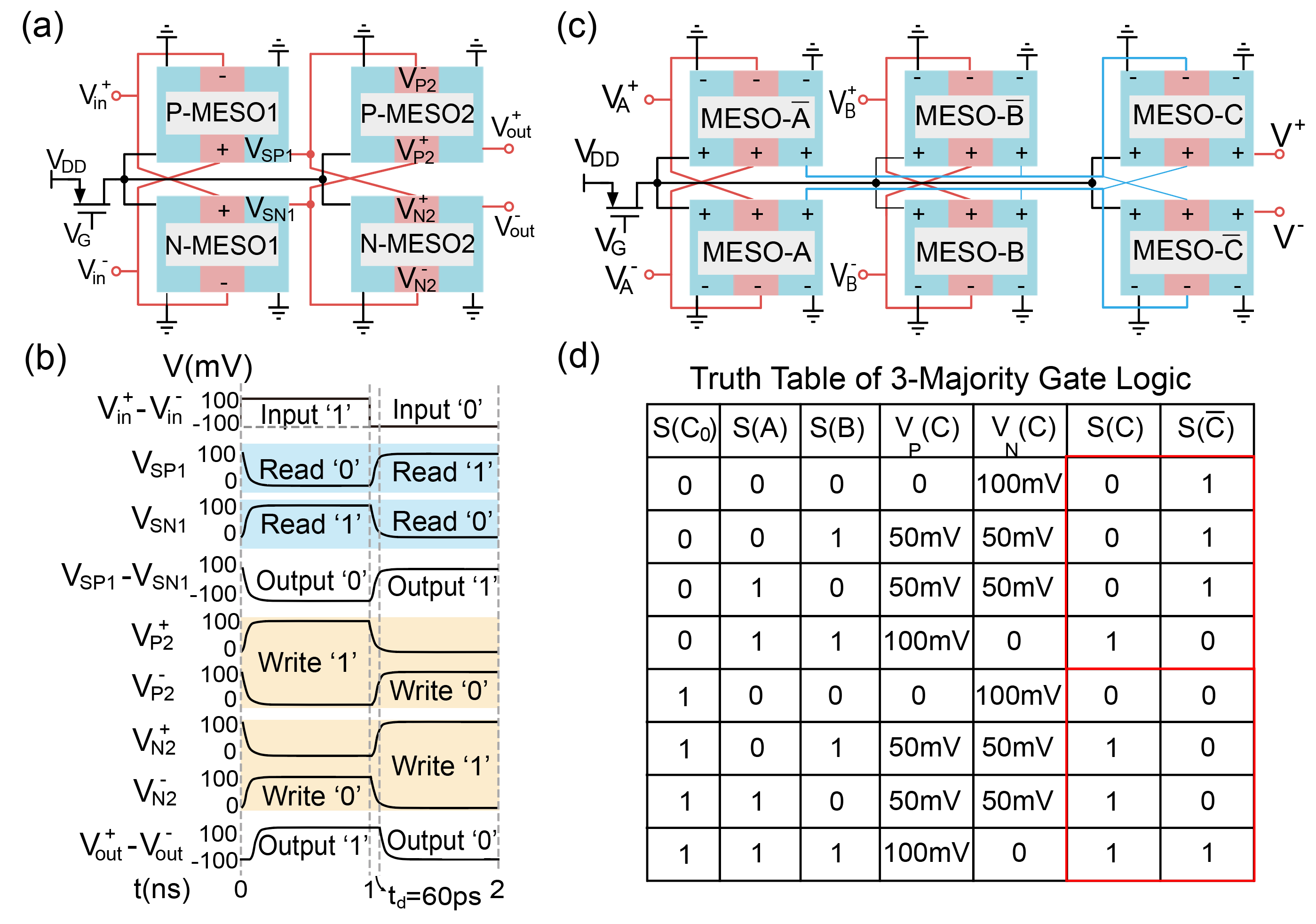}
    \caption{(a) Inverter chain design using C-MEMM. P-MEMMs are for holding the same bit as output, N-MEMMs are for holding the reversed bits. (b) Operating waveforms of the inverters, the propagation delay for one stage is 60ps.  (c) Majority gate logic, realizing logic-in-memory. (d) Truth table of applied voltages on cells A and B, and final states stored in cells C. 
    \vspace{-0.5cm}}
    \label{fig:logic}
\end{figure}

\begin{figure}[t]
    \centering
    \includegraphics[width=1\linewidth]{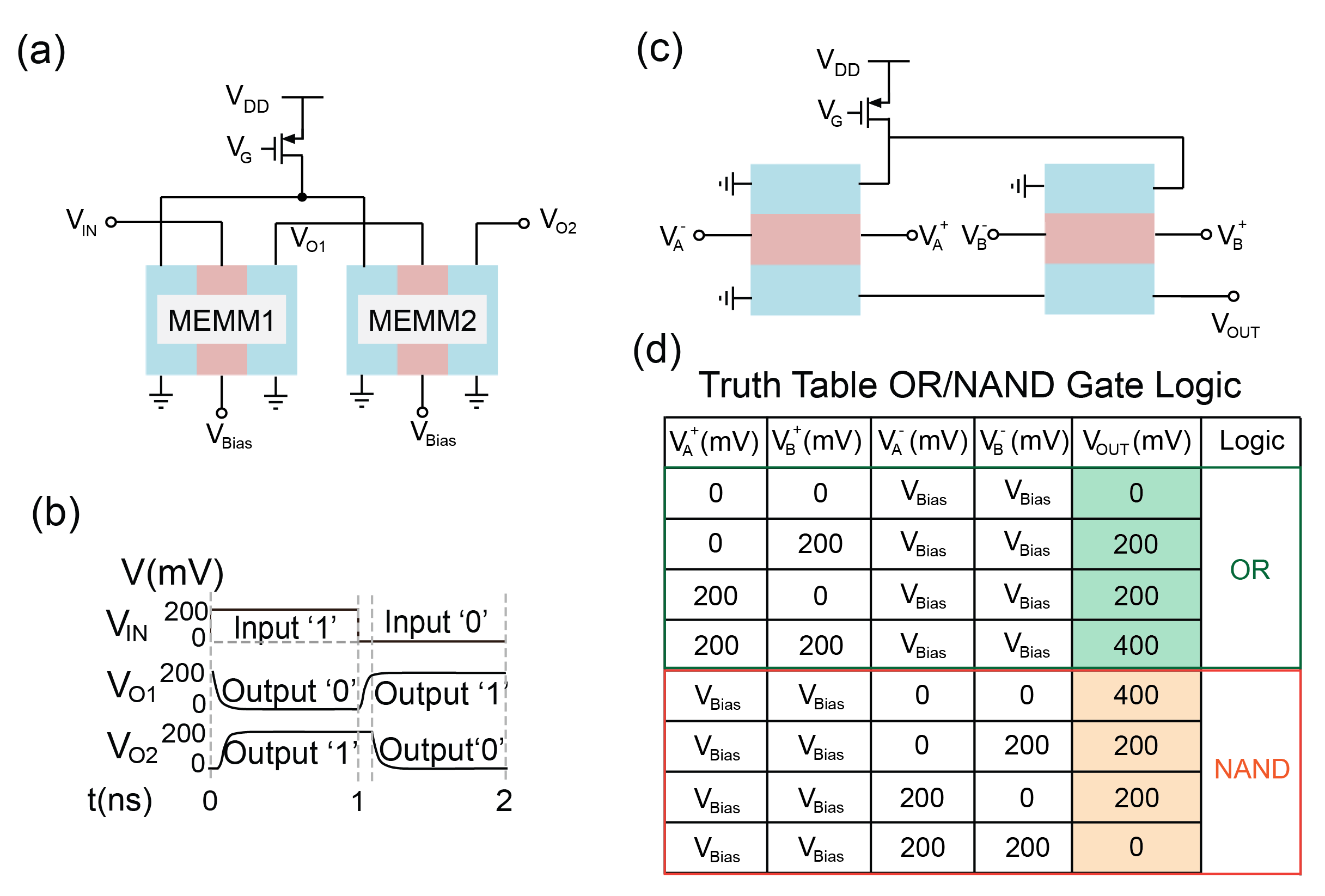}
    \caption{(a) Inverter chain design using single MEMM cells. Each cell stores the same bit as the output. (b) Operating waveforms, the propagation delay for one stage is 30ps. (c) OR/NAND gate logic. (d) Truth table of applied voltages on cells A and B, and output voltage at $V_\text{OUT}$.
    \vspace{-1.5em}}
    \label{fig:logic-singlecell}
\end{figure}

\vspace{-1.0em}
\subsection{Logic Gates Design}
\normalcolor

Unlike memory arrays, logic circuits benefit from a clear separation between the control (switching) path and the power (current) path. 
This is achieved by using MEMM cells with side ports, which allow direct switching of the ME layer without disturbing the main current path. 
The control path is indicated by the red switching lines in Fig.~\ref{fig:logic}(a).

\subsubsection{Complementary Logic Design}
Pairing two MEMM cells in a complementary configuration (C-MEMM) enables efficient implementation of various logic functions. This approach provides enhanced noise margins and lower power consumption, while producing a robust differential output voltage capable of directly driving subsequent logic stages.

Fig.~\ref{fig:logic}(a) depicts a two-stage inverter chain built from C-MEMM units. 
Each unit pairs a P-MEMM and an N-MEMM cell with opposite ferroelectric polarities: the P-MEMM defines its lower LBFO terminal as positive and generates a positive output when its polarization is up (P-up), while the N-MEMM defines its upper LBFO terminal as positive and outputs a positive voltage when its polarization is down (P-down). A switching voltage of 100mV is assumed for all devices.
The chain's operation is shown in Fig.~\ref{fig:logic}(b). 
Applying a differential input \(V^+_{\text{in}}-V^-_{\text{in}} = 100\)mV sets the first-stage outputs to \(V_{\text{SP1}} = 0\)mV (P-MEMM) and \(V_{\text{SN1}} = +100\)mV (N-MEMM), performing the first inversion. 
These complementary voltages drive the second stage, which inverts them again to produce \(V^+_{\text{out}} = +100\)mV and \(V^-_{\text{out}} = 0\)mV. 
A single shared power transistor supplies current to all C-MEMM cells in a logic block, minimizing area and transistor count.
The dynamic performance is also shown in Fig.~\ref{fig:logic}(b). 
Each C-MEMM stage exhibits a propagation delay of ~60ps, dictated primarily by interconnect resistance and the multiferroic capacitor charging time \cite{manipatruni_scalable_2019, parsonnet_toward_2020}. 
During a read, the differential output develops rapidly as the magnetic state modulates spin current. 
Switch operations, however, require a sufficiently long voltage pulse to overcome the ferroelectric coercive field and establish a stable polarization.

For implementing more complex logic, we propose a three-input majority gate (Fig.~\ref{fig:logic}(c)). Together with an inverter, this gate forms a functionally complete logic set. The circuit uses dual-rail C-MEMM cells for the inputs $A$, $B$, and $C$ and their complements ($\bar{A}$, $\bar{B}$, $\bar{C}$). Cells $A$ and $B$ are connected in parallel to generate a control voltage that drives the control port of the $C/\bar{C}$ cell pair.
We consider a high-state input cell as a Norton source $I_s$ in parallel with an internal resistance $R_s$, giving an output level $V_H = I_s R_s \approx 100\,\mathrm{mV}$. In the low state, the source is effectively off and only $R_s$ remains. The parallel combination of $A$ and $B$ therefore yields
\[
V_{AB}=
\begin{cases}
V_H, & A=B=1,\\[2pt]
V_H/2, & A\neq B,\\[2pt]
0, & A=B=0,
\end{cases}
\]
so only the agreement case reaches the switching threshold $V_{\mathrm{th}}\approx 100\,\mathrm{mV}$. When $A=B=1$, $V_{AB}\ge V_{\mathrm{th}}$ sets $C\rightarrow 1$. When $A=B=0$, the complementary network ($\bar{A}=\bar{B}=1$) analogously sets $\bar{C}\rightarrow 1$, i.e., $C\rightarrow 0$. If $A\neq B$, both rails remain below threshold and $C$ retains its prior state, which encodes the third input. Hence, the final state implements$Y=\mathrm{Maj}(A,B,C)=AB+AC+BC,$
as validated by the truth table in Fig.~\ref{fig:logic}(d). The stored output in $C/\bar{C}$ can directly drive subsequent stages, enabling cascaded majority logic.



\subsubsection{Single-Cell Logic Design}

For a simpler design, we also present a single-cell logic structure (Fig.~\ref{fig:logic-singlecell}). In the inverter configuration, a bias voltage $V_{\text{bias}}=100$~mV serves as a reference, while inputs are set to 0 or 200~mV. This scheme enables bidirectional electric fields that switch the cell between two logic states. The operating waveforms in Fig.~\ref{fig:logic-singlecell}(b) show both the inversion functionality and the temporal dynamics, with input transitions propagating through successive stages at 30ps delays.

This architecture also supports simplified logic gates. Fig.~\ref{fig:logic-singlecell}(c) shows a reconfigurable OR/NAND gate using two MEMM devices, where the logic function can be switched dynamically by the input configuration. The truth table in Fig.~\ref{fig:logic-singlecell}(d) maps input voltage pairs ($V_A^+$, $V_B^+$) to the output $V_{\text{OUT}}$. Green rows demonstrate OR logic (output high if either input is high), while orange rows show NAND logic (output low only when both inputs are high). Since NAND is functionally complete, this architecture can implement any Boolean function, offering a versatile platform for logic-in-memory computing.

\vspace{-1.0em}
\subsection{Deep Pipelining and Sequential Logic Design}

\begin{figure}[t]
    \centering
    \includegraphics[width=1\linewidth]{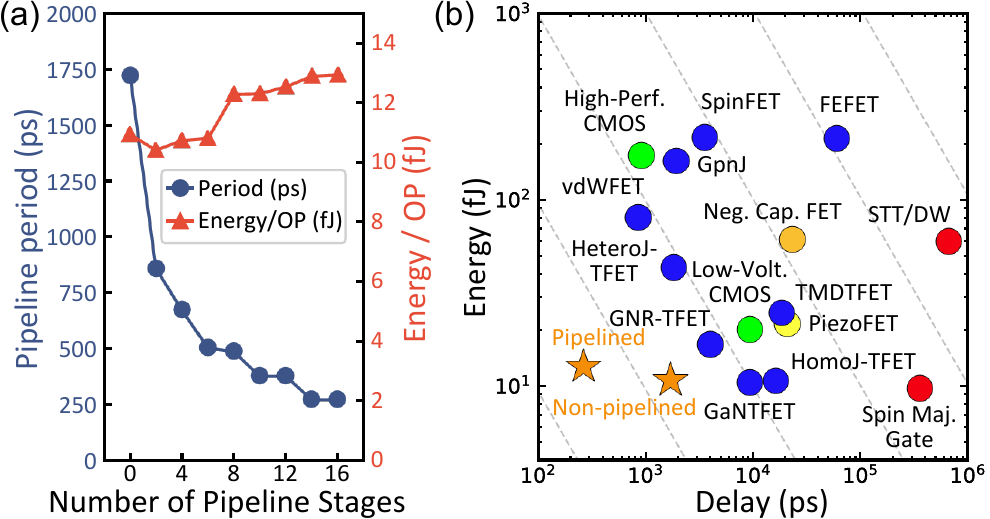}
    \caption{(a) The clock period and energy consumption of a 32-bit ALU implemented with MEMM cells. (b) Comparison of ALU performance with the same structure implemented using other devices \cite{nikonov_benchmarking_2015}.
    \vspace{-1.5em}}
    \label{fig:pipeline}
\end{figure}

To understand the potential of MEMM technology for sequential logic, consider a simple example: computing the Boolean function $F = ((A \cdot B) + (C \cdot D))$ for multiple input sets. In a conventional CMOS implementation, we would use two AND gates followed by an OR gate. To process multiple operations concurrently and increase throughput, designers employ pipelining, by inserting registers between logic stages to allow different inputs to be processed simultaneously at different stages. However, each register in CMOS requires multiple transistors (typically 6-12 per bit), consumes both dynamic and static power, and occupies significant silicon area. As pipeline depth increases, the overhead of these registers quickly dominates the circuit, limiting how aggressively designers can partition logic operations.


The logic-in-memory feature of MEMM devices fundamentally changes this trade-off. Because each MEMM cell is inherently non-volatile and stores its own state, every logic gate simultaneously acts as both a computational element and a storage register. There are no extra transistors, no additional area, and therefore, little energy overhead to insert pipeline stages. This means we can partition logic operations at an extremely fine granularity, placing a pipeline stage after every single logic gate if desired, without the penalties that make such aggressive pipelining impractical in CMOS.


To demonstrate this capability, we implemented a deeply pipelined 32-bit arithmetic logic unit (ALU), with the same topology and functionalities (bitwise NAND, NOR, addition, and subtraction) as in \cite{nikonov_benchmarking_2015}. We first designed a library of MEMM-based INV, NAND, AND, NOR, and OR gates, and characterized the delay, switching energy, and leakage power of each gate using the model described in Section III. 
{\color{black}More specifically, we set the supply voltage to be 200mV, and the leakage power to be 2.5nW and switching energy to be 1aJ}. 
We used Synopsys Design Compiler to construct the ALU netlist, automatically optimizing the pipeline stages with its register retiming features. To avoid timing conflicts, we controlled the pipelined ALU with non-overlapping clocks, i.e., consecutive pipeline stages alternate between computing and storing. 

As illustrated in Fig.~\ref{fig:pipeline}, the baseline ALU with no pipelining achieves a delay of 1722~ps while consuming 11.0~fJ per operation. It consists of a total of 1,576 MEMM cells. By partitioning the 32-bit ALU with pipelining and executing each pipeline stage concurrently, we can increase the throughput from 0.58 to 3.66~GOPS (giga operations per second), while the energy moderately increases to 12.9~fJ. This highlights the key distinction from CMOS, that MEMM cells directly store intermediate results between lock cycles, naturally forming the pipeline structure with no additional transistors.

The deep pipelining capability of MEMM technology opens up architectural possibilities beyond traditional von Neumann computing. By eliminating the distinction between computation and storage, MEMM circuits naturally support in-memory computing paradigms where data movement is minimized and computation occurs directly within the memory array. This spatial computing approach is particularly well-suited for data-intensive applications such as neural network inference, cryptographic operations, and signal processing, where the energy cost of data transfer often exceeds the energy required for computation itself. Future work will explore hierarchical memory architectures that leverage MEMM's unique properties to create energy-efficient accelerators for emerging workloads in artificial intelligence and edge computing domains.

\vspace{-1.0em}
\section{Conclusions}
We proposed and experimentally demonstrated a magnetoelectric magnon memory (MEMM) device that enables non-volatile, ultralow-power operation through electric-field-controlled switching and magnon-based signal transport. With circuit modeling, we demonstrated its scalability by designing both memory and logic arrays, showcasing MEMM’s potential for energy-efficient, in-memory computing architectures.

\vspace{-1.0em}
\section*{Acknowledgment}

R.Cong, R.Ramesh, and K.Yang acknowledge the support by NSF FUSE program (\#2329111). 
Z.Yao acknowledge the support by the U.S. Department of Energy, Office of Science, The Advanced Scientific Computing Research (ASCR) program, EXPRESS: 2025 Exploratory Research for Extreme-Scale Science, Analog Compute-in-Memory with Trainable Nonlinear Devices, and Microelectronics Science Research Center, under contract No. DE-AC02-05-CH11231.

\vspace{-1.0em}
\bibliographystyle{IEEEtran}
\bibliography{references}

\end{document}